# Dependence of Adhesive Friction on Surface Roughness and Elastic Modulus

Daniel Maksuta[1*], Siddhesh Dalvi[2*#], Abhijeet Gujrati[3], Lars Pastewka[4], Tevis DB Jacobs[4], Ali Dhinojwala[2@]

[1]Department of Biology, Integrated Bioscience Program, The University of Akron, Akron, Ohio, 44325, United States

[2]School of Polymer Science and Polymer Engineering, The University of Akron, Akron, Ohio 44325, United States

[3]Department of Mechanical Engineering and Materials Science, University of Pittsburgh, Pittsburgh, Pennsylvania, 15261, United States

[4]Department of Microsystems Engineering, University of Freiburg, 79110 Freiburg, Germany

#Present Address: Natural Fiber Welding, Peoria, IL, 61614, United States.

*Both authors contributed equally

@Corresponding Author: Ali Dhinojwala

**Email:** ali4@uakron.edu

**This PDF file includes:**

> Main Text
> Figures 1 to 5

**Abstract**

When adhesive elastomeric materials slide over hard rough surfaces at low velocities, there are two primary dissipative mechanisms that control how friction changes with sliding velocity: viscoelastic dissipation and adhesive dissipation. To distinguish the contribution of these dissipative mechanisms we have measured frictional shear stresses for crosslinked polydimethylsiloxane (PDMS) on three different rough surfaces of similar surface chemistry across nearly six decades of sliding velocity. The results show that the observed friction is dominated by adhesive dissipation, rather than viscoelastic dissipation. Prior models for elastomer friction assume that roughness only influences adhesive dissipation via the amount of contact area; by contrast, we find that the roughness-induced oscillations occurring across all length scales from macroscopic to atomic influence the molecular processes governing the adhesive dissipation. While it was previously known that roughness-induced oscillations affected the *viscoelastic* dissipation; this is the first demonstration that these oscillations also control the behavior of the *adhesive* component of friction. Finally, while theory predicts that rough friction should be independent of elastic modulus, we show that a strong dependence on modulus, with the frictional shear stress scaling as $E'^{1/2}$. When analyzed in this way, data from four different moduli and three different roughnesses collapse onto a universal curve to describe the velocity-dependent friction of soft materials. Taken together, this investigation sheds new light on the adhesive component of friction, and how it depends on the roughness and stiffness of the materials.

**Significance Statement**

Elastomer friction underlies critical applications from tires to soft robotics to 3D printing of soft materials. However, the scientific understanding is incomplete, with state-of-the-art models for rough surfaces describing only viscoelastic dissipation and neglecting the other major contribution: surface interactions. We fill this gap in understanding by performing friction experiments with varying-modulus silicone rubbers across a six-order-of-magnitude range of velocity, on surfaces that were characterized to atomic dimensions. Systematic shifts in velocity-dependent friction with roughness and stiffness suggest a universal curve of behavior for soft materials. These findings reveal how roughness-induced oscillations influence the interfacial interactions that control adhesive friction.

**Main Text**

**Introduction**

Soft adhesive materials, such as rubber, show frictional behavior markedly different than hard materials [1-4]. While the behavior at high sliding velocity or high stress can be described through macroscopic interfacial buckling instabilities, known as Schallamach waves [3-11], the behavior under smooth sliding is under active debate. At present, the smooth sliding of an elastomer on a rough surface is explained using two primary mechanisms of energy dissipation. The first dissipative process is often called "adhesive friction" [12,13,14] and arises due to the adhesion of the polymer chains, where the chains are stretched until they detach after which they reattach and the process repeats. The extent to which the polymer chains stretch is dependent on the stochastic nature of detachment; it is dependent on the adhesion strength, velocity, and temperature [13]. At slow sliding velocities the stochastic process dominates, causing the polymer chains to detach before substantial stretching. At high sliding velocities the average stretch of the chains increases, resulting in an increase of frictional forces. At even larger sliding velocities the shear forces decrease due to reduction in the number of chains reattaching back on the surface, effectively reducing the contact area. This chain adhesion model predicts an increase in friction with both elastic modulus and surface energy [12,13,14]. The second loss mechanism during sliding is viscoelastic dissipation due to the forced oscillations of the rubber at the interface when conforming to the roughness. Through these oscillations energy is dissipated via heat in proportion to the material's loss modulus and roughness, resulting in increased friction. In addition, the interfacial roughness also plays a key role in how the elastomer conforms to the surface and therefore influences the number of adhesive interactions and induced oscillations [2,15-22].

Currently there exists only one quantitative model (without any adjustable parameters) that relates a surface's roughness, measured over many length scales, to the viscoelastic dissipation during sliding [12-21]. While it has been shown that this model captures some of the trends observed for non-adhesive friction [12-19], the influence of roughness on the adhesive component to friction is underdeveloped. An empirical model has been proposed by Persson [20,21] to account for roughness using the chain-stretching model developed by Schallamach, Leonov and Chernyak [12,13]. In this model, adhesive friction depends on three quantities: the amount of contact area, the strength of the interactions, and how the strength and contact area change with sliding velocity. Persson proposes that roughness only influences the amount of contact area, without any influence on the strength of interaction, or how it changes with sliding velocity [12-14,20,21]. While conceptually this is reasonable, it neglects the possibility that the roughness-induced oscillations at the interface may influence the molecular processes that dictate the adhesion strength [2,23,24].

Here we have designed experiments to test how roughness influences these dissipative mechanisms. We have chosen three surfaces that vary in roughness but with similar surface chemistry (non-polar bonds) that are described in detail in Refs. [25,26,27]. This should ensure that any differences in the observed friction should be primarily due to differences in surface roughness. We have monitored the contact area and cantilever deflection using high-speed video for almost 6 decades of sliding velocity for elastomers (PDMS) that vary in elastic modulus from 0.7 to 10 MPa while keeping the normal load constant. In this paper we discuss the contributions to friction from both the viscoelastic and adhesive processes occurring. Note that these experiments also shed light on the influence of roughness on Schallamach wave instabilities observed at high velocities, but this will be discussed in a follow up publication.

**Results and Discussion**

The power-spectral densities (PSDs) of the three surfaces [25,26,27] are shown in Figure 1a and the corresponding maximum shear force as a function of sliding velocity in Figure 1b. The smoothest surface is labeled as OTS and is an octadecyltrichlorosilane-coated single-crystal silicon wafer, followed by the slightly rougher pUNCD (polished ultrananocrystalline diamond) and the roughest UNCD (ultrananocrystalline diamond). The frictional shear stress during sliding is calculated using deflection and contact area for each frame as obtained from high-speed video of the contact (see Methods). The onset sliding velocity for Schallamach waves, $V_{SW-onset}$, was obtained by visual inspection of the high-speed video and is noted with the vertical dashed lines in Figure 1b. We find that $V_{SW-onset}$ decreases with increasing surface roughness, but the onset stress is independent of surface roughness. Once $V_{SW-onset}$ for the rough surfaces is achieved, a stress plateau regime occurs over two decades of velocity [3]. The end of the plateau regime coincides with the onset of Schallamach waves for OTS. After $V_{SW-onset}$ for smoothest OTS surface, the frictional shear stresses of all surfaces overlap and hence become independent of surface properties [3]. Based on these results, surface roughness only plays a dominant role in the pre-Schallamach wave regime.

To deduce in what way the differences in surface roughness are influencing the observed frictional shear stress, we first calculate the viscoelastic component, $\sigma_{visc}$, using Persson's model for friction on rough surfaces. This model is summarized in four equations below [15-21]:

1. $\sigma_{visc} = \int_{q_0}^{q_1} dq q^3 C(q) S(q, V_{sliding}) P(q, V_{sliding}) \int_0^{2\pi} d\phi \cos\phi \frac{E''(qV_{sliding}(t)\cos\phi, T_0)}{(1-v^2)}$
2. $S(q, V_{sliding}) = \gamma + (1-\gamma) P(q, V_{sliding})^2$
3. $P(q, V_{sliding}) = \text{erf}\left(\frac{1}{2\sqrt{G}}\right)$
4. $G(q, V_{sliding}) = \frac{1}{8} \int_{q_0}^{q} dq q^3 C(q) \int_0^{2\pi} d\phi \left|\frac{E'(qV_{sliding}(t)\cos\phi, T_0)}{(1-v^2)\sigma_0}\right|^2$

Here, $E'(qV_{sliding}(t)\cos\phi, T_0)$ and $E''(qV_{sliding}(t)\cos\phi, T_0)$ are the storage and loss modulus respectively evaluated at frequency $qV_{sliding}(t)\cos\phi$ and temperature $T_0$, $\phi$ is the orientation of the surface roughness relative to the sliding direction, $S(q, V_{sliding})$ is a correction factor for the size of the deformation zone, $P(q, V_{sliding})$ is the ratio of actual contact area to the projected contact area, $v$ is the Poisson ratio, $\gamma = 1/2$ is a numerical factor (see Ref. [21]), and $\sigma_0$ is the normal stress. Note that the expression $\int_{q_0}^{q_1} dq\, q^3 C(q)$ is the aspect-ratio of surface roughness squared or mean-square slope. Since in conformal contact the slope is essentially the strain introduced into the material, Eq. (1) can be roughly interpreted as multiplying the strain at scale $q$ with the loss modulus. The model predicts that roughness can either increase or decrease the magnitude of the observed frictional shear stress or cause the velocity at which friction is maximum to shift. To utilize Equations 1-4 requires denoting the bounds of integration $q_0$ to $q_1$, where $q_0$ is inversely proportional to the contact length and $q_1$ is inversely proportional to the length scale at which wear processes occur [28]—which is generally determined empirically based on the quality of the fit.

It is also important to note that the conformality predicted using Equations 2-4 neglects adhesive forces [15-21]. While this assumption may be valid for very hard rubbers and or very high sliding velocities, the calculated Tabor parameter for our system (assuming 1 to 10 nm asperity radius and surface energy of 40 mJ/m$^2$) and the low sliding velocity range suggests we are in the JKR contact limit where adhesion will be important [29]. From this consideration it is unclear if the calculated conformality is valid; thus, instead we calculate the upper and lower bounds for the viscoelastic component by assuming either completely conformal contact ($P(q, V_{sliding}) = 1$) or non-adhesive contact, respectively. Where the non-adhesive contact, otherwise known as Hertzian contact, is assumed for Equations 3 and 4. Figure 2 shows the calculated bounds for the viscoelastic component using the storage and loss modulus from Ref. [24]. While there is ambiguity in the magnitude of the viscoelastic component, ranging from 100 times smaller for the lower bound to comparable for the upper bound, it can be observed that regardless of the conformality the calculated stress for UNCD is greater than pUNCD and OTS across all sliding velocities. Even though the predicted trend of UNCD>pUNCD>OTS matches the experimental results, there are clear quantitative differences between the theoretical predictions and experiment. This suggests that the dominating mechanism for friction must be related to adhesion—which is not unexpected considering the low loss moduli for PDMS, and that adhesive friction is expected to dominate at low sliding velocities [2,20,21].

The theory for adhesive friction was developed by Schallamach, and later modified by Chernyak and Leonov et al. [12,13]. Persson incorporated the influence of roughness by accounting for regions of non-contact via the pre-factor $P(q_1, V_{sliding})$ and simplified the velocity dependence by using an exponential function [20,21]. The equation is

$$5.\ \sigma_{adh} = P(q_1, V_{sliding}) \left\{ \tau_{f0} \exp\left(-c \left[\log_{10}\left(\frac{V_{sliding}}{V_0}\right)\right]^2\right) \right\},$$

where $\sigma_{adh}$ is the frictional shear stress, $\tau_{f0}$ is the interfacial strength at vanishing sliding velocity, $c$ is a non-dimensional constant related to the rate of increase of frictional shear stress with sliding velocity, $V_0$ is the velocity at which $\sigma_{adh}$ is maximum, and $P(q_1, V_{sliding})$ is the area ratio at length-scale $q_1$. In this equation roughness influences the adhesive response via $P(q_1, V_{sliding})$, i.e. roughness only influences the amount of contact area. The term in the curly bracket in Equation 5 is a result of stretching of chains attached to the surface, and the parameters $\tau_{f0}$, $c$ and $V_0$ are expected to be independent of roughness.

We first consider if the differences in friction could be caused by differences in the real contact area for adhesive interactions, which is related to both the conformality via the prefactor $P(q_1, V_{sliding})$ in Eq. (5), and the total amount of surface area available. However, based on differences in conformality, OTS should have a larger adhesive stress than pUNCD and UNCD by a factor of 4 and 16, respectively, which is not observed. If the stress is proportional to the differences in available surface area than the ratio of surface areas of pUNCD and UNCD relative to OTS, 1.01 and 1.69, respectively, should correspond to the ratio of frictional shear stresses [26]. However, this is clearly not represented in the data either since experimentally UNCD and pUNCD are approximately 12 and 8 times larger than OTS, respectively. This suggests that the variation in surface roughness, instead of influencing the magnitude of the adhesive response, may be primarily influencing the velocity at which the adhesive component is maximum, i.e. the parameter $V_0$ that shows up in Eq. (5). This is not expected given the premises of the Schallamach or Persson molecular theory.

To test the hypothesis that roughness may influence $V_0$, we introduce the dimensionless sliding velocity $V_{sliding}^* = V_{sliding}/V_{SW-onset}$, where $V_{SW-onset}$ is the onset velocity for Schallamach waves for the pUNCD, UNCD, and OTS surfaces. Plotting the frictional shear stress versus $V_{sliding}^*$ collapses the data to a master curve (Fig. 3a), indicating that $V_{SW-onset}$ takes the role of $V_0$ in Eq. (5). The Schallamach molecular theory assumes that dissipation is driven by periodic excitation of surface chains. For conformal contact, the amplitude of excitation at frequency (see Figure 4) $\omega = qV_{SW-onset}V_{sliding}^* \equiv qV_{sliding}$ is related to the PSD $C(q)$. Indeed, plotting the PSD versus $qV_{SW-onset}$ (Figure 3b) collapses pUNCD and UNCD at low frequencies and OTS and pUNCD at high frequencies; this is reasonable agreement considering that $V_{SW-onset}$ was estimated visually from high-speed video over a sparsely populated set of $V_{sliding}$ and likely carries a large uncertainty in addition to the error associated with the PSDs.

These observations suggest that surfaces sliding at the same $V_{sliding}^*$ experience the same excitation spectrum and hence produce the same frictional shear stress. Since the normalizing velocity $V_{SW-onset}$ is the onset of Schallamach waves, this further indicates that Schallamach waves occur at some critical excitation spectrum that is identical for the three surfaces. The excellent collapse of the frictional shear stress can only be rationalized if there is little difference in the real contact area between the three rough surfaces with the same modulus. For nonadhesive contact, OTS should have approximately 4 and 16 times more contact area than pUNCD and UNCD, respectively. We assume that the presence of adhesive interactions plays a key role in influencing the extent of this conformality. Lastly, while there is an apparent velocity shift to the data with little difference between the magnitude of friction after shifting, the presence of the stress plateau region (correlated with the emergence of Schallamach waves) for the rough surfaces precludes our ability to determine if the magnitude of the maximum frictional shear stress is also influenced by roughness or if it is purely a velocity shift. The fact that the whole excitation spectrum seems to matter implies that length scales greater than the molecular length scale are important in determining the adhesive frictional response [12,13,14] and that the roughness-induced oscillations ($qV_{sliding}$) influence the velocity dependence for both the viscoelastic and adhesive component similarly.

Finally, we measured the effect on friction with variation in elastic modulus of the PDMS. While there is no expectation for the modulus dependence of $V_{SW-onset}$, based on Equation 5 and for surfaces of similar chemistry, we expect friction to be independent of elastic modulus. This is because the increase in friction due to increased elastic modulus via $\tau_{f0}$ should be offset by the loss of contact area via $P(q_1, V_{sliding})$—i.e. $P(q_1, V_{sliding})$ scales as $1/E'$ (Figure S1, SI) while $\tau_{f0}$ scales as $E'$, thus the product should be independent of elastic modulus. Additionally, so long as the loss tangent for the material is independent of elastic modulus the variation of $P(q_1, V_{sliding})$ with $V_{sliding}$ will be independent of elastic modulus—which is assumed for the calculations in Figure S1, SI. What we find that $V_{\text{SW-onset}}$ is largely independent of elastic modulus (where $V_{\text{SW-onset}}$ occurs between two adjacent experimental sliding velocities across all moduli), and that the friction within a modulus collapses to form a master curve when normalizing by $V_{\text{SW-onset}}$; suggesting that there is little difference in the contact area within a modulus. However, we observe a trend that friction is not independent of elastic modulus (Figure 5a), suggesting that the proposed scaling of Equation 5 does not capture the observed trends. Instead, we find that after normalizing the data by $E'^{1/2}$, the results collapse for these four moduli of PDMS elastomers ranging from 0.7 MPa to 10 MPa (Figure 5b). The frictional shear stress data were modeled using a linear regression (using a log-log plot) after normalizing with $E'^x$, where the $x$ was varied between 0 and 1.5. The $R^2$ values obtained from the fits are shown in Figure S2 (SI) and we find best fits for $x$ between 0.5-0.75, consistent with the collapse of the data in Figure 5B after scaling the frictional shear stress by $E'^{1/2}$. Interestingly, frictional shear stress scaling as $E^{\frac{1}{2}}$ was also proposed by Chaudhury [15] based on the simple model proposed by Ludema and Tabor and is consistent with what we are observing here [30]. This result suggests that the presence of surface forces mitigates the loss of contact area with increasing elastic modulus and/or strengthens the interfacial interactions more than expected. Both possibilities can be rationalized when considering how adhesion increases with increasing elastic modulus, summarized by the Tabor parameter [29].

Even though Equation 5 provides a good conceptual framework for understanding the influence of roughness on friction in the regime where adhesion plays an important role, both the experimental observation of velocity shift and dependence on modulus do not match the predictions from Equations 3-5. Thus, we are unable to deconvolute the contributions to friction that are attributed to the number of interactions and their corresponding strength for rough surfaces. Future theoretical work needs to determine the impact of adhesion on $P(q_1, V_{sliding})$, or conformality, in addition to roughness, modulus, and velocity [31]. Accurate predictions of $P(q_1, V_{sliding})$ will provide a more complete understanding of friction in the smooth-sliding regime for soft elastomers on rough surfaces.

**Conclusion**

For PDMS of various elastic moduli sliding on three rough surfaces with similar interfacial chemistry we find that interfacial roughness plays a key role in influencing the friction behavior in the smooth-sliding regime. The friction here is found to be dominated by adhesive dissipation rather than viscoelastic dissipation. To account for the influence of roughness on adhesion, the molecular adhesion model proposed by Schallamach has been modified by introducing a pre-factor that accounts for changes in contact area due to roughness, elastic modulus, and velocity. Our results demonstrate two main factors that play an important role in quantitatively explaining the influence of roughness and modulus on friction. First, we find that the adhesive response is shifted in sliding velocity in proportion to the variation in roughness, suggesting that the velocity dependence of the adhesive response is influenced by a range of length scales rather than just the molecular length scale, as previously thought [12,13,14]. Thus, our results reveal a previously unknown design parameter to control friction in the smooth sliding regime. Second, contrary to the expectation from theory that friction on rough surfaces should be independent of elastic modulus, we find that frictional shear stress scales with $E'^{1/2}$. This result suggests that the presence of surface forces mitigates the loss of contact area with increasing elastic modulus and/or strengthens the interfacial interactions more than expected—both of which can be explained by considering how adhesion varies with elastic modulus via the Tabor parameter [30]. Overall, this investigation provides the critical data to elucidate the behavior of adhesive friction, and its dependence on the roughness and stiffness of the materials.

**Materials and Methods**

The friction measurements for the range of sliding velocities were done using a nano-stepper motor (NewFocus) for slow velocities (nm/s to $\mu$m/s) and a Servo Motor (Moog Animatics SM) with different pitch sizes for faster velocities ($\mu$m/s to m/s). A normal load of 5 mN was used for each experimental trial. The shear force was measured using a double-cantilever spring on which a PDMS lens was attached (supplemental Figure S3). The cantilever deflection, $L$, and contact area, $A$, were measured in each frame of the high-speed video (60 to 30,000 frames per second) recorded using a Photron FASTCAM SA-04 mounted to an Olympus microscope. The cantilever deflection was measured by tracking sharp detectable edges in the video using Pro-analyst (Xcitex) software and the contact area was measured by using an edge detection script in MATLAB (Mathworks). The frictional shear force, $F$, was calculated by multiplying the deflection by the spring constant, $k$: $F = kL$. The frictional shear stress, $\sigma_s$, was calculated by normalizing the frictional shear force and contact area per fame: $\sigma_s = F/A$. The spring calibration curve and spring constant are shown in supplemental Figure S4 is calculated from three repeats. The method for preparation of polydimethylsiloxane (PDMS) hemispherical lenses and the diamond rough surfaces is described elsewhere [25,26,27]. The OTS surfaces were characterized using stylus profilometer and AFM. Data was stitched together via averaging in the region of shared wavenumber $q$.

**Acknowledgments**

We thank Edward Laughlin, the machinist at the University of Akron, for building the in-house friction setup. We thank Prof. BNJ Persson for useful discussions. A.D. acknowledges funding from the NSF (DMR-1610483). TDBJ acknowledges support from the NSF (CMMI-1727378).

**Figures**

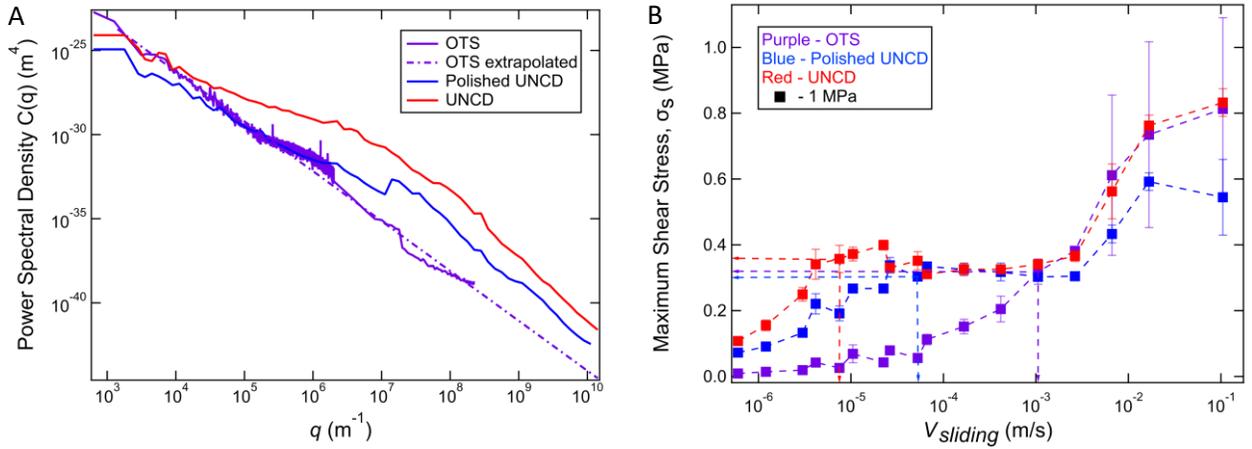

**Figure 1.** *Topography characterization of the three surfaces and the corresponding frictional shear stresses. A) PSD for the three surfaces measured from millimeter to Ångstrom length scales. The value of $C(q)$ (Power Spectral Density) which is proportional to the amplitude of the roughness increases from OTS (purple) to pUNCD (blue) to UNCD (red). We use extrapolation for OTS to estimate the roughness in the nanometer to Ångstrom length scales (dotted purple line). B) Example of experimentally observed frictional shear stress versus sliding velocity, $V_{sliding}$, for the three surfaces (OTS, pUNCD, and UNCD) with PDMS of 1 MPa modulus. The vertical dashed lines correspond to the onset velocity for Schallamach waves (buckling instability), $V_{SW-onset}$, observed for the different surfaces. We will limit our analysis to the pre-buckling instability regime. Normal force is $5\ mN$. Error bars correspond to standard deviations calculated from three repeats.*

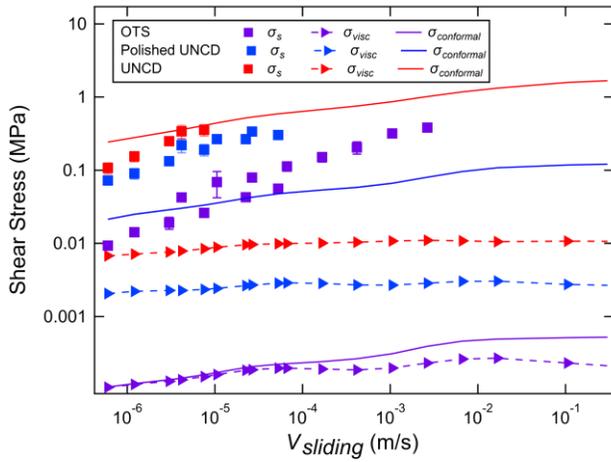

**Figure 2.** *Log-log plot of the calculated viscoelastic component assuming conformal contact (solid lines, $\sigma_{\text{conformal}}$) and Hertzian contact (dashed lines with triangles, $\sigma_{\text{visc}}$) plotted with experimental frictional shear stress data for 1 MPa PDMS plotted against sliding velocity, $V_{sliding}$, in the pre-instability regime. Experimental data is plotted as square symbols. Calculated viscoelastic contribution does not conform to the trends observed in the experimental data regardless of expected extent of conformality. Storage and loss moduli data comes from Ref. [24]. Error bars correspond to standard deviations calculated from three repeats.*

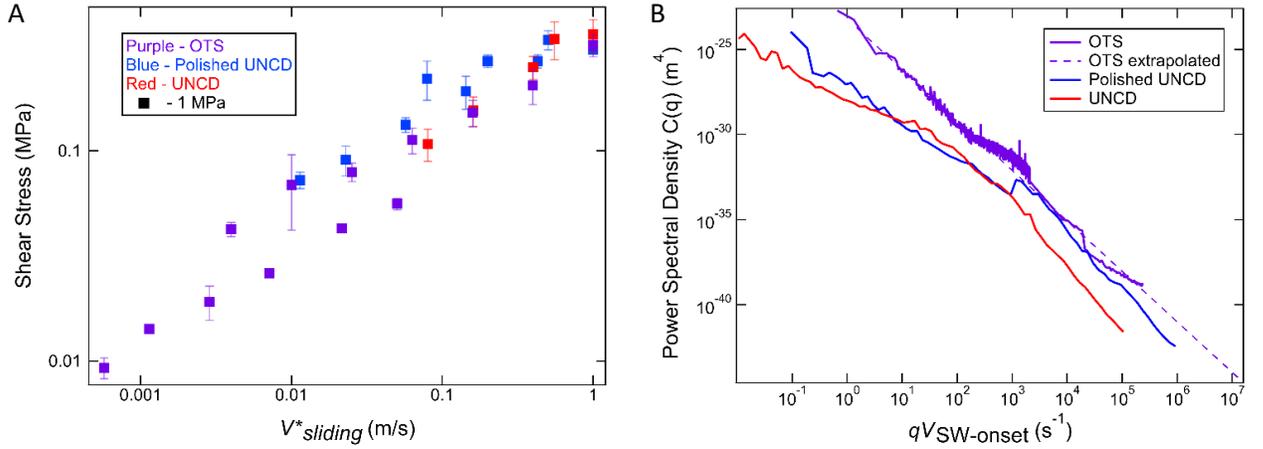

**Figure 3.** *Experimental frictional shear stresses and PSDs collapse after x-axis are shifted with respect to $V_{SW-onset}$. A) Log-log plot of frictional shear stress for 1 MPa PDMS plotted versus $V^*_{sliding} = V_{sliding}/V_{SW-onset}$. B) $C(q)$ plotted versus $qV_{SW-onset}$, where $V_{SW-onset}$ is the onset sliding velocity for Schallamach waves, measured from the high-speed videos. $V_{SW-onset}$ for the various surfaces: $V_{UNCD} = 7.5 \times 10^{-6}$ m/s, $V_{pUNCD} = 6.62 \times 10^{-5}$ m/s, $V_{OTS} = 1.05 \times 10^{-3}$ m/s. Normal force is 5 mN. Error bars correspond to standard deviations calculated from three repeats.*

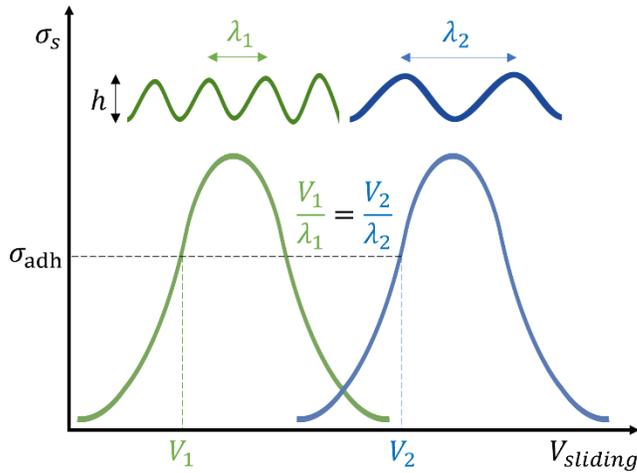

**Figure 4.** *A sketch representing the physical mechanism behind the observed velocity shift of the adhesive component to friction. Adhesive shear stress, $\sigma_{adh}$, is constant when the oscillation frequency, $V_{sliding}/\lambda$ (where $\lambda = 2\pi/q$), and magnitude of roughness, h, are equal. The adhesive shear stress is shifted to lower sliding velocities as the wavelength decreases from $\lambda_2$ to $\lambda_1$. Note that, while similar oscillation-frequency arguments have been previously used to describe viscoelastic friction, these are the first results that demonstrate that it also describes adhesive friction.*

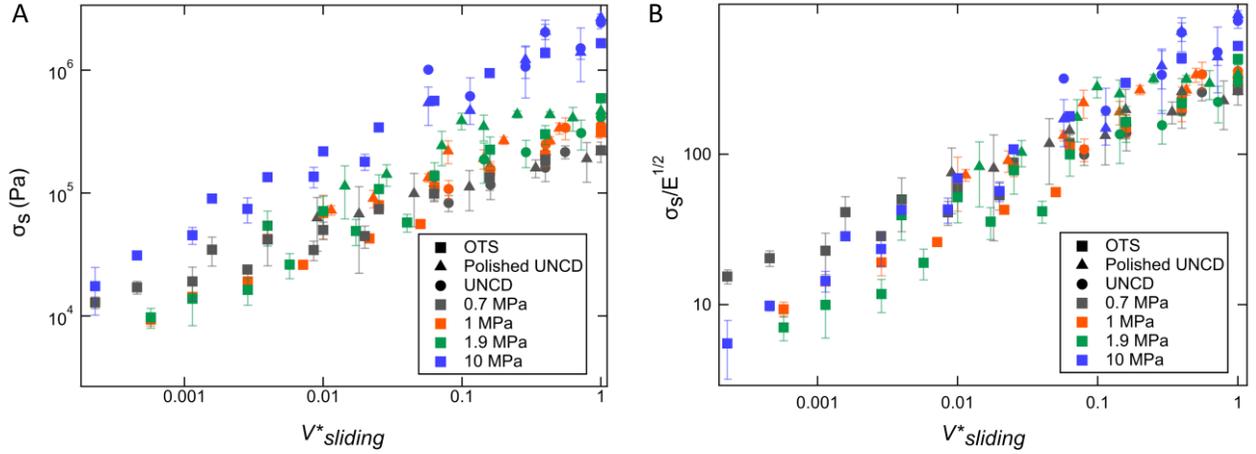

**Figure 5.** *Frictional shear stress as a function of elastic modulus collapses to form a master curve when normalized by elastic modulus raised to the ½ power. A) Log-Log plot of frictional shear stress versus $V^*_{sliding}$ for 0.7, 1, 1.9, and 10 MPa PDMS in the pre-instability regime. Expectation from Equation 5 is that friction should be independent of elastic modulus. B) Log-Log plot of frictional shear stress, $\sigma_s$, normalized by the respective elastic modulus raised to the ½ power, plotted against $V^*_{sliding}$ in the pre-instability regime. The overlap of the data after normalization suggests the pre-factor to the adhesive contribution, which is a product of the magnitude and number of interactions, should scale as $\sqrt{N/m^2}$. Normal force is 5 mN. Error bars correspond to standard deviations calculated from three repeats.*

# Supporting Information to "Dependence of Adhesive Friction on Surface Roughness and Elastic Modulus"

Daniel Maksuta[1*], Siddhesh Dalvi[2*#], Abhijeet Gujrati[3], Lars Pastewka[4], Tevis DB Jacobs[4], Ali Dhinojwala[2]

[1]Department of Biology, Integrated Bioscience Program, The University of Akron, Akron, Ohio, 44325, United States

[2]School of Polymer Science and Polymer Engineering, The University of Akron, Akron, Ohio 44325, United States

[3]Department of Mechanical Engineering and Materials Science, University of Pittsburgh, Pittsburgh, Pennsylvania, 15261, United States

[4]Department of Microsystems Engineering, University of Freiberg, 79110 Freiberg, Germany

#Present Address: Natural Fiber Welding, Peoria, IL, 61614, United States.

*Both authors contributed equally to this work

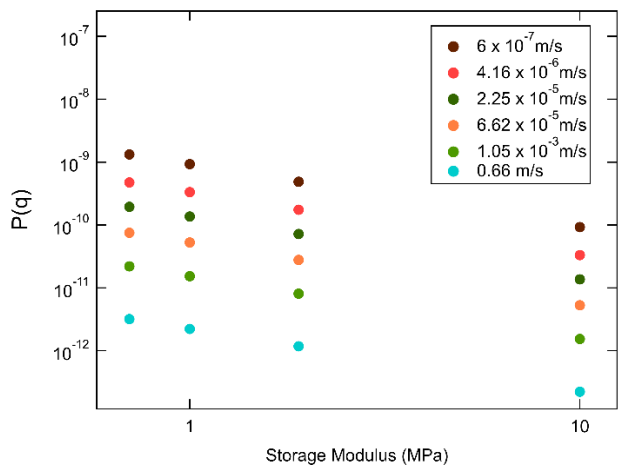

**Figure S1.** *Calculated P(q) for four different moduli across a range of velocities. The slope of this graph implies that P(q) scales as 1/E.*

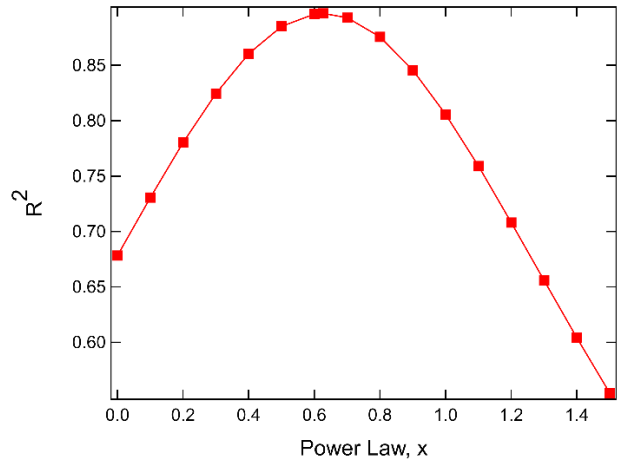

**Figure S2**. *calculated R^2 for figure versus modulus normalization exponent, $E'^{x}$, for figure 5b. Normalization is optimal at approximately 0.6274 with one standard deviation giving an upper and lower bound of about 0.5 and 0.75.*

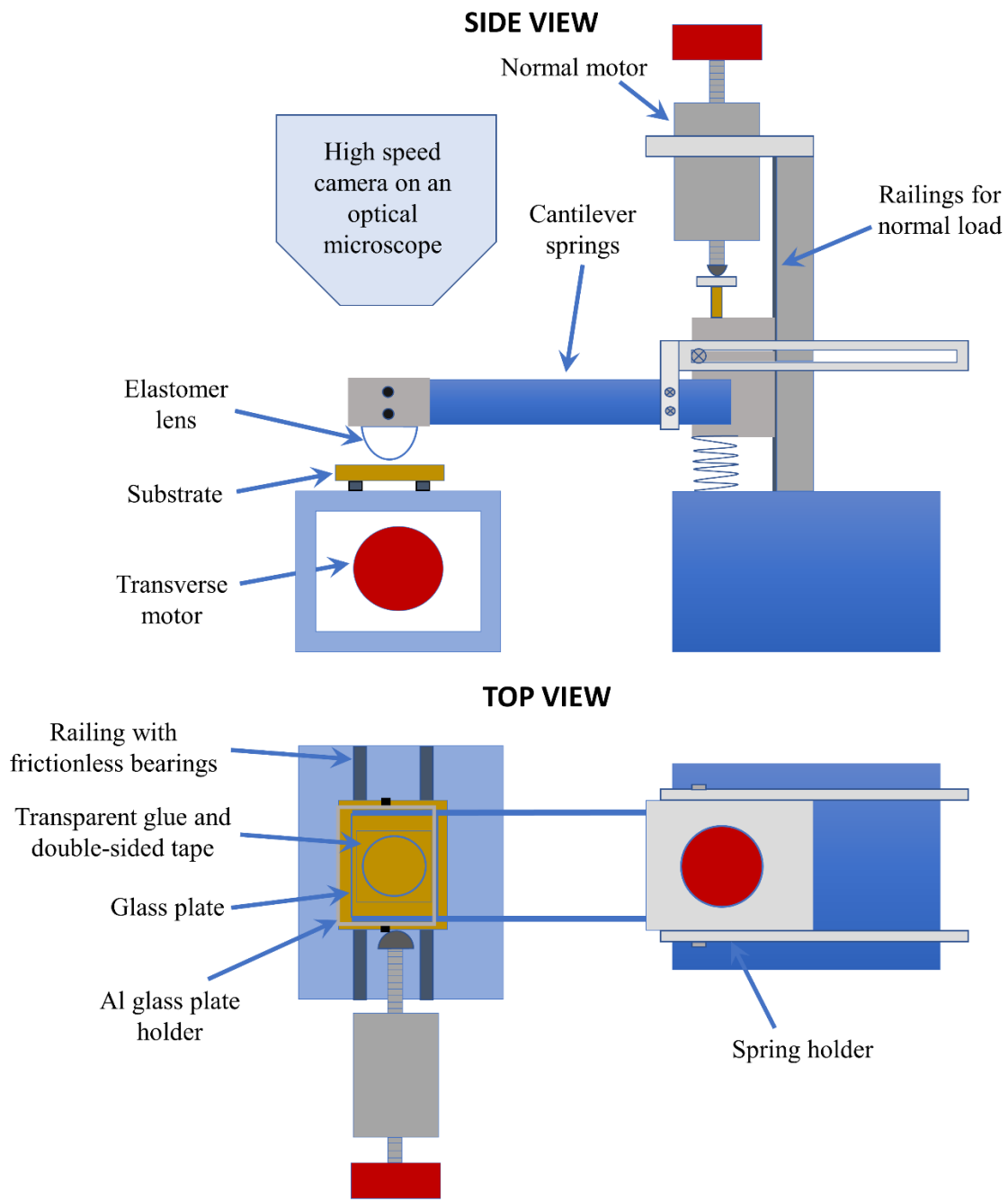

**Figure S3.** *The double-cantilever setup for friction measurement through spring deflection and real-time area tracking through high-speed camera inspired by refs. [1,2].*

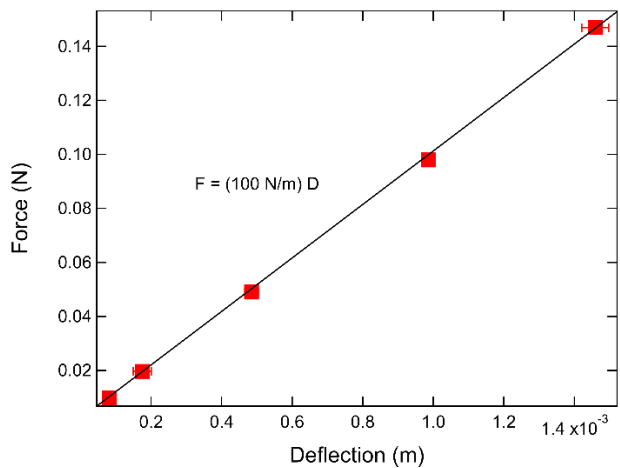

**Figure S4.** *Spring constant calibration. Force was applied to the spring and the corresponding deflection was measured.*

**References**

1. S. P. Arnold, A. D. Roberts, and A. D. Taylor, Rubber Friction Dependence on Roughness and Surface Energy, *J. Nat. Rubb. Res.*, **2**, 1, (1987).

2. K. Vorvolakos and M. K. Chaudhury, The Effects of Molecular Weight and Temperature on the Kinetic Friction of Silicone Rubbers, *Langmuir*, **19**, 6778 (2003).